\documentclass[aps,prl,twocolumn,floatfix,superscriptaddress,showpacs,amsfonts,amssymb,amsmath,preprintnumbers]{revtex4}
\usepackage{bm}
\usepackage{graphicx}
\usepackage{color}

\renewcommand{\eqref}[1]{Eq.\ (\ref{#1})}

\newcommand{\eqsand}[2]{Eqs.\ (\ref{#1}) and (\ref{#2})}

\newcommand{\Eqsand}[2]{Eqs.\ (\ref{#1}) and (\ref{#2})}
\newcommand{\figref}[1]{Fig.\ \ref{#1}}

\newcommand{\bea}{\begin{eqnarray}}
\newcommand{\eea}{\end{eqnarray}}
\newcommand{\beq}{\begin{equation}}
\newcommand{\eeq}{\end{equation}}
\newcommand{\lt}{\left}
\newcommand{\rt}{\right}
\newcommand{\bl}{\bigl}
\newcommand{\br}{\bigr}
\newcommand{\la}{\langle}
\newcommand{\ra}{\rangle}

\newcommand{\mbf}[1]{\bm{#1}}

\newcommand{\dd}{\partial}
\newcommand{\vdel}{\boldsymbol{\nabla}}
\newcommand{\vdperp}{\vdel_\perp}
\newcommand{\dperp}{\nabla_\perp}

\newcommand{\vz}{\hat{\mbf{z}}}
\newcommand{\vk}{\mbf{k}}

\newcommand{\kperp}{k_\perp}
\newcommand{\kpar}{k_\parallel}
\newcommand{\kparf}{k_{\parallel \rm f}}
\newcommand{\kf}{k_{\perp \rm f}}
\newcommand{\kbox}{k_{\perp \rm box}}
\newcommand{\ko}{k_0}
\newcommand{\km}{k_m}
\newcommand{\kc}{k_c}
\newcommand{\vA}{v_A}
\newcommand{\omA}{\omega_A}
\newcommand{\uperp}{u_\perp}
\newcommand{\tA}{\tau_A}
\newcommand{\tnl}{\tau_{\rm nl}}
\newcommand{\tnlo}{\tau_{\rm nl,0}}
\newcommand{\tnlkf}{\tau_{\rm nl,1}}

\newcommand{\vu}{\mbf{u}}
\newcommand{\vuperp}{\vu_\perp}
\newcommand{\vB}{\mbf{B}}
\newcommand{\dvBperp}{\delta\vB_\perp}

\newcommand{\zk}{\zeta_{\kpar}}
\newcommand{\zo}{\zeta_0}
\newcommand{\zkf}{\zeta_1}
\newcommand{\zkn}{\zeta_n}
\newcommand{\Ekn}{E_n}
\newcommand{\Eo}{E_0}
\newcommand{\Mo}{M_0}
\newcommand{\Ekf}{E_1}
\newcommand{\Fpm}{F^\pm}

\newcommand{\eps}{\varepsilon}

\begin{document}


\title{Weak Alfv\'en-Wave Turbulence Revisited}
\author{Alexander~A.~Schekochihin}
\email{a.schekochihin1@physics.ox.ac.uk}
\affiliation{Rudolf Peierls Centre for Theoretical Physics, University of Oxford, Oxford OX1 3NP, United Kingdom}
\author{Sergey~V.~Nazarenko}
\affiliation{Mathematics Institute, University of Warwick, Coventry CV4 7AL, United Kingdom}
\author{Tarek~A.~Yousef}
\affiliation{Bj{\o}rn Stallars gt 1, 7042 Trondheim, Norway}
\date{\today}
 
\begin{abstract}
Weak Alfv\'enic turbulence in a periodic domain is considered as a mixed state of 
Alfv\'en waves interacting with the two-dimensional (2D) condensate. 
Unlike in standard treatments, no spectral continuity between the two is assumed 
and indeed none is found. If the 2D modes are not directly forced, 
$k^{-2}$ and $k^{-1}$ spectra are found for the Alfv\'en waves and the 2D 
modes, respectively, with the latter less energetic than the former. 
The wave number at which their energies become 
comparable marks the transition to strong turbulence. 
For imbalanced energy injection, the spectra are similar and the Elsasser 
ratio scales as the ratio of the energy fluxes in the counterpropagting Alfv\'en waves. 
If the 2D modes are forced, a 2D inverse cascade dominates the dynamics at the largest scales, 
but at small enough scales, the same weak and then strong regimes as 
described above are achieved. 
\end{abstract}


\maketitle

\paragraph{Introduction.} It has been understood for many years that small-scale 
turbulence of a conducting fluid or plasma in a strong magnetic field consists of 
Alfv\'en-wave packets \cite{Kraichnan}. 
This is true in most astrophysical plasmas, including the weakly collisional 
ones, where Alfv\'enic fluctuations populate the scales 
above the ion Larmor scale \cite{GK2}. 
On theoretical \cite{Montgomery_Turner,GS95,GS97}, 
numerical \cite{Shebalin_etal,Cho_Vishniac,Maron_Goldreich,Mueller_etal} 
and observational \cite{Bieber_etal,Bigazzi_etal,SorrisoValvo_etal,Horbury_etal,Wicks_etal,Chen_Mallet} 
grounds, it appears clear that Alfv\'enic turbulence is anisotropic 
with $\kperp\gg\kpar$. 
The parallel scales are associated with the propagation of Alfv\'en waves, 
the perpendicular ones with the nonlinear interaction between them. 
The relative importance of these effects depends on 
the corresponding time scales, $\tA\sim(\kpar v_A)^{-1}$ 
and $\tnl\sim(\kperp\uperp)^{-1}$, where 
$v_A$ is the Alfv\'en speed and $\uperp$ the perpendicular velocity 
perturbation. 
When $\tA\gg\tnl$, the nonlinearity dominates and the turbulence is effectively 
two-dimensional (2D); 
when $\tA\ll\tnl$, the wave propagation dominates and the turbulence is weak. 

A causality argument suggests that a pure 2D regime cannot be sustained: 
for any given $\kperp$, motions in two planes perpendicular 
to the mean field and separated by a distance $\sim\kpar^{-1}$ can only 
remain correlated if the time it takes an Alfv\'en wave to propagate 
between the planes is $\tA<\tnl$. Thus, an initially 2D perturbation will 
naturally decay into a state of ``critical balance,'' $\tA\sim\tnl$. 
It has been argued \cite{GS97,Galtier_etal,Lithwick_Goldreich} that weak 
turbulence, the limit case opposite to 2D, will also approach critical balance 
via a perpendicular cascade in which $\tnl$ becomes ever smaller until 
$\tnl\sim\tA$ at some sufficiently small scale. 
The critical balance thus appears to be the fundamental physical 
principle underpinning Alfv\'enic turbulence 
(and possibly in general turbulence in systems that support 
propagation of waves \cite{Nazarenko_Schekochihin}). However, the structure of 
critically balanced turbulence remains contentious and poorly understood. 
Efforts to improve this understanding have often turned to 
various insights from the theory of weak turbulence 
for intuition and guidance. Weak Alfv\'en-wave turbulence itself, while 
having for some time enjoyed the reputation of a solved problem \cite{Galtier_etal}, 
has nevertheless recently been subject of several new investigations that 
amended or disagreed with the 
established paradigm \cite{Nazarenko_enslave,Boldyrev_Perez_condensate,Wang_Boldyrev_Perez,Alexakis}. 
This paper is a contribution to this revisionist tendency, focusing 
on the structure of weak Alfv\'en-wave turbulence in finite periodic 
domains and on the transition to critical balance. 
While a periodic box may be an artificial setting, it is ubiquitous in numerical 
experiments. It is, therefore, important to understand turbulence in such 
domains and the extent to which it might belong to the same 
universality class as turbulence in natural systems.

While it might appear that weak turbulence is an analytically tractable and, therefore,
easily understood limit, the weak turbulence of Alfv\'en waves is, in fact, 
difficult to treat in a rigorous fashion because of a special role played 
by the modes with $\kpar=0$. Since Alfv\'en waves have frequencies 
$\omega^\pm_{\vk}=\pm\kpar v_A$ and only counterpropagating waves can interact,
the resonance conditions $\omega^+_{\vk_1}+\omega^-_{\vk_2}=\omega^\pm_{\vk_3}$ and 
$\vk_1+\vk_2=\vk_3$ imply that at least one mode in any interacting triad must 
have $\kpar=0$ \cite{Montgomery_Matthaeus,Ng_Bhattacharjee}. 
These modes are not Alfv\'en waves but rather 2D motions for which 
$\tA=\infty$, so they cannot be treated by the weak-turbulence approximation. 
The standard remedy for this complication has been to proceed 
with the weak-turbulence expansion anyway, assuming formally that the 
$\kpar$ spectrum of the Alfv\'en waves is continuous across $\kpar=0$
\cite{Galtier_etal,Lithwick_Goldreich}. The resulting theory predicts 
a $\kperp^{-2}$ scaling of the energy spectrum. As this scaling 
is corroborated by direct numerical simulations \cite{Perez_Boldyrev_weak}, 
the continuity assumption might seem to be 
vindicated. In this paper, we propose a different way of treating the 
$\kpar=0$ modes, with no assumption of their spectral continuity with the Alfv\'en waves.
It leads to a new phenomenological theory of the weak Alfv\'en-wave turbulence 
and to distinct scaling predictions for the energy spectra of the Alfv\'en waves 
and of the $\kpar=0$ modes. We discuss various regimes: balanced and 
imbalanced, containing forced 2D motions or otherwise, and also describe 
the transition to critically balanced strong turbulence in a new way. 
 
\paragraph{Scaling theory.} Let us start with the equations of reduced magnetohydrodynamics 
(RMHD) \cite{Kadomtsev_Pogutse,Strauss},
which can be shown to describe correctly the anisotropic Alfv\'enic fluctuations 
both in an MHD fluid \cite{Montgomery}, 
and, above the ion Larmor scale, even in weakly collisional kinetic plasmas \cite{GK2}. 
In RMHD, the velocity and magnetic field perturbations perpendicular 
to the mean field $\vB_0=\vA\vz$ (the Alfv\'enic perturbations) 
are two-dimensionally solenoidal, so they can be 
expressed in terms of stream and flux functions: 
$\vuperp = \vz\times\vdperp\Phi$ and $\dvBperp = \vz\times\vdperp\Psi$.  
The RMHD equations can then be written in terms of the Elsasser potentials 
$\zeta^\pm = \Phi\pm\Psi$ as follows \cite{GK2}
\beq
\label{RMHD}
\dd_t\zeta^\pm\mp\vA\dd_z\zeta^\pm = 
N[\zeta^\mp,\zeta^\pm] + \Fpm,
\eeq
where $\Fpm$ is the stream function of a body force representing energy injection 
and the nonlinear term~is
\begin{align}
\nonumber
N[\zeta^\mp,\zeta^\pm] = -{1\over2}\,\dperp^{-2}&\lt(\lt\{\zeta^+,\dperp^2\zeta^-\rt\} 
+ \lt\{\zeta^-,\dperp^2\zeta^+\rt\}\rt.\\ 
&\lt.\mp \dperp^2\lt\{\zeta^+,\zeta^-\rt\}\rt),
\end{align}
where $\{A,B\}\equiv\vz\cdot(\vdperp A\times\vdperp B)$.

Let us now Fourier transform in $z$, 
factor out the oscillating in time part of the solution,
$\zeta^\pm(x,y,z) = \sum_{\kpar}\zk^\pm(x,y)e^{\pm i\kpar\vA t + i\kpar z}$,
and write separately the evolution equations for the Alfv\'en waves ($\kpar\neq0$), 
\begin{align}
\nonumber
\dd_t\zk^\pm = N[\zo^\mp,\zk^\pm] 
+ N[\zk^\mp,\zo^\pm]e^{\mp i2\kpar\vA t}\qquad\quad\\
+ \sum_{\kpar'\neq0,\,\kpar} N[\zeta_{\kpar'}^\mp,\zeta_{\kpar-\kpar'}^\pm]e^{\mp i2\kpar'\vA t} 
+ \Fpm_{\kpar} e^{\mp i\kpar\vA t},
\label{eq_AW}
\end{align}  
and the $\kpar=0$ modes,
\beq
\dd_t\zo^\pm = N[\zo^\mp,\zo^\pm] 
+ \sum_{\kpar\neq0} N[\zk^\mp,\zeta_{-\kpar}^\pm] e^{\mp i2\kpar\vA t}.
\label{eq_zmodes}
\eeq

The standard approximation of the weak-turbulence theory is, roughly speaking, 
to neglect the nonlinear terms in \eqref{eq_AW} that have oscillatory factors, 
so the dominant effect is the Alfv\'en-wave ``scattering'' off the 
$\kpar=0$ modes (the first term on the right-hand side). 
This gives rise to a cascade of energy to small perpendicular scales 
(large $\kperp$), while the transfer of energy from 
the directly forced $\kpar$ to other $\kpar$ is small. 
For simplicity, let us assume that only one $\kpar=\kparf$ is forced. 
If we denote $\zkn^\pm(\kperp)$ the characteristic amplitudes 
corresponding to $\kpar=n\kparf$ and the perpendicular wave number $\kperp$ 
and take the interactions to be local in $\kperp$, 
we may estimate
\beq
(\dperp^2\zkf^\pm)N[\zo^\mp,\zkf]\sim
\kperp^4\zo^\mp(\kperp)[\zkf^\pm(\kperp)]^2\sim\eps^\pm,
\label{adv_AW}
\eeq
where $\eps^\pm=\la(\vdperp\zeta^\pm)\cdot(\vdperp\Fpm)\ra$ is the mean power injected 
by the forcing into the ``$+$'' and ``$-$'' modes. 
We first consider the {\em balanced} case: $\eps^+=\eps^-=\eps$ 
\footnote{Physically speaking, this 
is the only case we are allowed to model by \eqref{RMHD} because we 
must in fact set $F^+=F^-$ in order to ensure that only velocity 
field is directly forced. Indeed, if $F^+-F^-\neq0$, we would be introducing 
an inhomogeneous forcing term in the induction equation, leading to unphysical 
breaking of the magnetic flux conservation at the forcing scale. 
This said, $F^+\neq F^-$ is a popular modeling choice for simulating 
imbalanced MHD turbulence and we will consider this case later on.}. 

\eqref{adv_AW} shows that in order to make scaling predictions for 
the Alfv\'en waves, we must know the scaling of the amplitudes 
of the $\kpar=0$ modes --- these modes determine the 
cascade rate $\sim\kperp^2\zo^\mp(\kperp)$ for the Alfv\'en waves. 
The $\kpar=0$ modes are very different from the Alfv\'en waves: they are described 
by 2D magnetohydrodynamics with an oscillatory nonlinear source term 
representing the coupling of $\kpar$ and $-\kpar$ Alfv\'en waves 
[\eqref{eq_zmodes}]. We shall first consider the situation in which 
the $\kpar=0$ modes are not forced externally, so this nonlinear source is
the only source of energy in the $\kpar=0$ modes. Since the nonlinear source in 
\eqref{eq_zmodes} has an oscillatory factor, there is a strong cancellation 
effect and the amplitude of the $\kpar=0$ modes can be estimated as
\footnote{A reader who does not find this estimate obvious on dimensional 
grounds, may be convinced by the following argument. 
Consider a model stochastic equation for the $\kpar=0$ modes: 
$\dd_t \zo = -\zo/\tau_c + \chi(t)e^{-i\omA t}$, where $\tau_c\sim\tnlo$ is the decorrelation time 
due to nonlinear mixing of $\zo$ by and the $\chi$ term models the oscillatory coupling to 
the Alfv\'en waves [see \eqref{eq_zmodes}].  
The forcing amplitude $\chi(t)\sim\kperp^2\zkf^+\zkf^-$ also has 
decorrelation time $\tau_c$ because $\zkf^\pm$ are also mixed by $\zo$ 
[see \eqref{eq_AW}]. 
Taking $\la\chi(t)\chi^*(t')\ra = \la|\chi|^2\ra e^{-|t-t'|/\tau_c}$ 
and solving our model equation, we obtain 
$\la|\zo|^2\ra\sim\la|\chi|^2\ra/\omA^2$, q.e.d.} 
\beq
\zo^\pm(\kperp) \sim \omA^{-1}\kperp^2\zkf^+(\kperp)\zkf^-(\kperp), 
\label{amp_zmodes}
\eeq
where $\omA=\kparf\vA$. 
Combining \eqsand{adv_AW}{amp_zmodes}, we find 
\begin{align}
\label{formula_E1}
&\zkf^\pm(\kperp) \sim 
(\eps\omA)^{1/4}\kperp^{-3/2}
\Rightarrow 
\Ekf^\pm(\kperp) \sim 
(\eps\omA)^{1/2}\kperp^{-2}, \\
&\zo^\pm(\kperp) \sim 
\lt(\frac{\eps}{\omA}\rt)^{1/2}\kperp^{-1}
\Rightarrow 
\Eo^\pm(\kperp) \sim 
\frac{\eps}{\omA}\,\kperp^{-1}, 
\label{formula_E0}
\end{align}
where $\Ekn^\pm(\kperp)$ is the one-dimensional energy spectrum, 
related to the characteristic amplitudes via 
$\kperp \Ekn^\pm(\kperp) \sim \kperp^2[\zkn^\pm(\kperp)]^2$.
The energy injection is balanced, so the the ``$+$'' and ``$-$'' spectra 
have the same scaling. 

The phenomenological argument presented above has led to a prediction 
of the Alfv\'en-wave spectrum [\eqref{formula_E1}] 
that is formally the same as the prediction 
of the standard weak-turbulence theory (this is only true 
for the balanced case; see below). However, the physical origin of this 
spectrum is different and the assumption of continuity across $\kpar=0$ is certainly 
not satisfied:  
\beq
\frac{\Eo^\pm(\kperp)}{\Ekf^\pm(\kperp)} \sim \frac{\tA}{\tnlo} 
\sim \lt(\frac{\tA}{\tnlkf}\rt)^2 \sim \frac{\kperp}{\kc} \ll1,
\label{Eo_ratio}
\eeq
where $\tA=\omA^{-1}$, $\tnlo\sim[\kperp^2\zo^\pm(\kperp)]^{-1}$,
$\tnlkf\sim[\kperp^2\zkf^\pm(\kperp)]^{-1}$, 
and $\kc=(\omA^3/\eps)^{1/2}$. 
The amplitudes of the $\kpar=0$ modes are smaller than those of the Alfv\'en waves 
as long as $\tA\ll\tnl$, i.e., as long as the weak-turbulence limit holds. However, 
their spectrum is shallower than that of the Alfv\'en waves, so the ratio between 
the amplitudes increases with $\kperp$ until they become comparable at $\kperp\sim\kc$. 
At this point the turbulence becomes critically balanced and is no longer weak, 
developing a $\kperp^{-5/3}$ \cite{GS95,Beresnyak} 
or perhaps $\kperp^{-3/2}$ \cite{Boldyrev_align} spectrum.

\begin{figure}[t]
\centerline{\includegraphics[width=8cm]{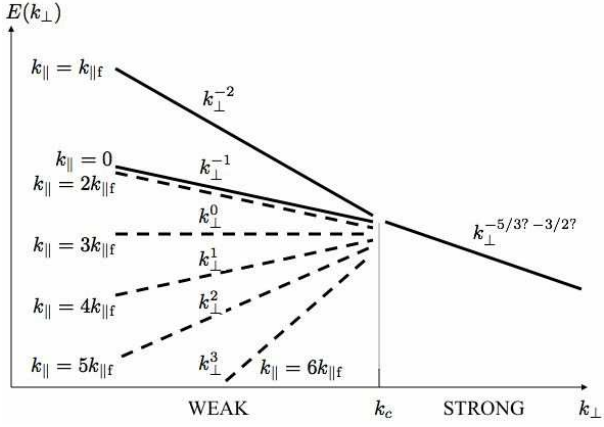}}
\caption{\label{fig_spectra_cartoon} A schematic illustration 
of the weak-turbulence spectra and transition to strong, critically 
balanced turbulence.}
\end{figure}

\paragraph{Higher-$\kpar$ modes.} 
In the same way that a small leakage of energy from the forced modes ($\kpar=\kparf$) 
via oscillatory nonlinear couplings gives rise to a spectrum of  
$\kpar=0$ modes, similar leakages arise from $\kpar=\kparf$ to $\kpar=2\kparf$ 
and from there onwards to $\kpar=3\kparf$, $4\kparf$, \dots, $n\kparf$ 
(the third term on the right-hand of \eqref{eq_AW}). 
These modes do not contribute to the scattering of the driven modes 
and are just induced by the latter. Analogously to \eqref{amp_zmodes}, 
it is not hard to see that for $n\ge2$,
\beq
\zkn^\pm(\kperp) \sim {\kperp^{2(n-1)}[\zkf^\pm(\kperp)]^n\over\omA^{n-1}},
\eeq
whence follow the spectra of these modes [using \eqref{formula_E1}]
\beq
\Ekn(\kperp)\sim {\kperp^{3(n-1)}\over\omA^{2(n-1)}}\,\Ekf^n(\kperp)
\sim {\eps^{n/2}\over\omA^{(3n-4)/2}}\,\kperp^{n-3}.
\eeq
Similarly to \eqref{Eo_ratio}, we have 
$\Ekn(\kperp)/\Ekf(\kperp)\sim\lt(\kperp/\kc\rt)^{n-1}$,
so the amplitudes of the $\kpar=n\kparf$ modes with $n\ge2$ become comparable to the amplitude 
of the forced mode ($n=1$) at the same wave number as does the amplitude of the $\kpar=0$  
mode. This situation is illustrated schematically in \figref{fig_spectra_cartoon}.

\paragraph{Imbalanced turbulence.} If the energy injection into the ``$+$'' and ``$-$'' 
Elsasser modes is not the same, say, $\eps^+>\eps^-$, 
the resulting turbulence is known as {\em imbalanced} --- this is, in fact, a generic 
situation in the solar wind \cite{Tu_etal,Lucek_Balogh_imb} and also in numerical simulations 
if one considers local subdomains of the simulation box \cite{Perez_Boldyrev_patches}. 

Our arguments are easily adapted to this case. 
\Eqsand{adv_AW}{amp_zmodes} hold unchanged (note that the latter 
formula implies $\zo^+\sim\zo^-$). \Eqsand{formula_E1}{formula_E0} 
generalize to 
\begin{align}
\label{formula_E1_imb}
&\zkf^\pm(\kperp) \sim 
\frac{(\eps^\pm)^{3/8}}{(\eps^\mp)^{1/8}}\frac{\omA^{1/4}}{\kperp^{3/2}}
\Rightarrow 
\Ekf^\pm(\kperp) \sim 
\omA^{1/2}\frac{(\eps^\pm)^{3/4}}{(\eps^\mp)^{1/4}}\,\kperp^{-2}, \\
&\zo^\pm(\kperp) \sim 
\frac{(\eps^+\eps^-)^{1/4}}{\omA^{1/2}\kperp}
\Rightarrow 
\Eo^\pm(\kperp) \sim 
\frac{(\eps^+\eps^-)^{1/2}}{\omA}\,\kperp^{-1}, 
\label{formula_E0_imb}
\end{align}
These imply that the cross-helicity and the Elsasser ratio in weak 
imbalanced turbulence are idependent of $\kperp$: 
\beq
\label{amps_vs_fluxes}
\Eo^+(\kperp)\sim\Eo^-(\kperp),\qquad
\frac{\Ekf^+(\kperp)}{\Ekf^-(\kperp)} \sim \frac{\eps^+}{\eps^-}, 
\eeq

Note that the standard weak-turbulence treatment \cite{GS97} of the imbalanced case 
was insufficient to fix the individual scalings of the ``$+$'' and ``$-$'' 
spectra \cite{Lithwick_Goldreich}. Our argument does not have this problem 
and is able predict the relationship between amplitudes and fluxes [\eqref{amps_vs_fluxes}],
for which the standard weak-turbulence theory had to make recourse to 
solving the kinetic equation, formally invalid for the $\kpar=0$ modes 
\cite{Galtier_etal}. 

The $\kpar=0$ modes are still small compared to the Alfv\'enic modes: 
\beq
\frac{\Eo^\pm(\kperp)}{\Ekf^\pm(\kperp)} \sim \lt(\frac{\tA}{\tnlkf^\pm}\rt)^2 
\sim \frac{\kperp}{\kc^\pm} \ll1,
\label{Eo_ratio_imb}
\eeq
where the nonlinear times $\tnlkf^\pm\sim[\kperp^2\zkf^\mp(\kperp)]^{-1}$ 
are now different for the ``$+$'' and ``$-$'' modes and so are the 
wave numbers $\kc^\pm=\omA^{3/2}(\eps^\pm)^{1/4}(\eps^\mp)^{-3/4}$ 
at which the weak-turbulence approximation breaks down. 
It must of course break down already at the smaller of the two, viz., 
$\kc^-$ (we have assumed $\eps^+>\eps^-$). 
This opens the possibility of a ``twighlight range'' of $\kperp$ 
between $\kc^-$ and $\kc^+$ (or another threshold dependent 
on the intermediate scalings) --- an extended transition from weak to strong 
regime, where the ``$-$'' modes are strongly nonlinear, while the ``$+$'' modes 
are not. It is indeed possible to construct such mixed theories, featuring 
steeper spectra for the ``$+$'' modes and shallower ones for the 
``$-$'' modes (cf.\ \cite{Chandran_imb}). 
We are tempted to speculate that this might help explain the origin 
of apparently non-universal (and different) slopes of the ``$+$'' and ``$-$'' 
spectra found in simulations of strongly imbalanced Alfv\'enic turbulence
\cite{Beresnyak_imb,Mallet_imb}. However, the more or less 
arbitrary assumptions necessary to fix scalings in this regime and the consequent 
uncertainties in the outcome are so numerous that we prefer not to treat this 
subject here. The precise scalings of strong turbulence in the imbalanced regime 
also remain theoretically uncertain 
\cite{Lithwick_imb,Beresnyak_imb,Chandran_imb,Perez_Boldyrev_patches,Podesta_imb}. 

\paragraph{Case of hydrodynamically forced $\kpar=0$ modes.} We have so far considered a special 
case in which the $\kpar=0$ modes were not forced. 
Let us now allow comparable power to be injected into $\kpar=0$ as into $\kpar=\kparf$. 
The situation is now radically different because the amplitude of the 
$\kpar=0$ modes is no longer determined by a small leakage of the 
Alfv\'en-wave energy via the oscillatory term in \eqref{eq_zmodes}, but 
by a direct forcing. If we ignore the oscillatory term altogether 
(assuming it averages out to lowest order in $\tA/\tnl$), the $\kpar=0$ 
modes decouple and form an independent 2D turbulent condensate. We 
write the equations for this condensate in terms of 
its velocities, given by the stream function 
$\Phi_0 = (\zeta_0^+ + \zeta_0^-)/2$, and 
magnetic fields, given by the flux function 
$\Psi_0 = (\zeta_0^+ - \zeta_0^-)/2$: 
\begin{align}
\label{Phi0_eq}
&\dd_t\dperp^2\Phi_0 + \lt\{\Phi_0,\dperp^2\Phi_0\rt\} = \lt\{\Psi_0,\dperp^2\Psi_0\rt\} 
+ \dperp^2 F_0,\\
&\dd_t \Psi_0 + \lt\{\Phi_0,\Psi_0\rt\} =
{\rm Re} \sum_{\kpar\neq0} \bl\{\zk^+,\zeta_{-\kpar}^-\br\} e^{- i2\kpar\vA t}, 
\label{Psi0_eq}
\end{align}
where we have returned to the assumption that all forcing is in the velocity field.
Therefore, the magnetic flux function $\Psi_0$ 
is a passive scalar and its only source of energy  
is the oscillatory coupling to the Alfv\'en waves, which has
consequently been retained in \eqref{Psi0_eq}. 
Since this energy source is small, $\Psi_0\ll\Phi_0$ and the Lorentz 
force can be neglected in \eqref{Phi0_eq}, leaving a 2D Euler equation. 
For the forced Alfv\'en waves with $\kpar = \kparf$, we have from from 
\eqref{eq_AW}, neglecting the oscillatory terms and $\Psi_0$,
\beq
\dd_t\zeta_1^\pm = N[\Phi_0,\zeta_1^\pm] + F_1 e^{\mp i\kpar\vA t}.
\label{eq_forced}
\eeq
The unforced Alfv\'en waves with $\kpar\neq\kparf$ will have small 
amplitudes due to the oscillatory terms in \eqref{eq_AW}. 

\begin{figure}[t]
\centerline{\includegraphics[width=8cm]{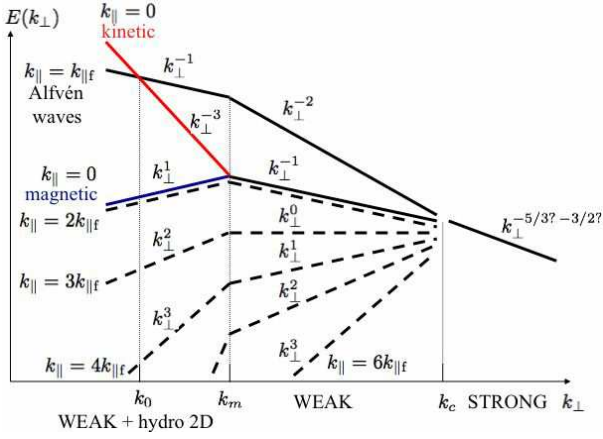}}
\caption{\label{fig_invcasc} A schematic illustration 
of the weak-turbulence spectra for the case of velocity-forced 
$\kpar=0$ modes.}
\end{figure}

Thus, the turbulence has split into the following distinct components: 
a 2D ($\kpar=0$) forced {\it hydrodynamical} condensate, 
an ensemble of forced Alfv\'en waves passively advected by this 2D condensate 
[\eqref{eq_forced}], 
small 2D magnetic fluctuations feeding off the Alfv\'en waves and 
also passively advected by the 2D hydrodynamic condensate 
[\eqref{Psi0_eq}], 
and small unforced Alfv\'en waves again passively advected 
by the 2D condensate and feeding off the forced modes and each other 
(in a way similar to that described above for the case of no forcing of 
the $\kpar=0$ modes). This situation is somewhat similar to the 
``slaved'' regime proposed in \cite{Nazarenko_enslave}, where everything 
is unilaterally controlled by the 2D condensate. 

Since the condensate is hydrodynamical and 2D, one expects an inverse 
energy cascade to perpendicular scales larger than the forcing scale
(in a finite system leading to energy accumulation at the system scale) 
\footnote{An inverse cascade in weak MHD turbulence has recently 
been reported \cite{Alexakis}. The regime with an inverse cascade and 
spectra similar to \figref{fig_invcasc} 
was also found by T.~A.~Yousef in unpublished numerical work (2008).}. 
Below the forcing scale, the direct enstrophy cascade produces a well-known 
kinetic-energy spectrum: 
\beq
\Eo(\kperp)\sim\kperp[\Phi_0(\kperp)]^2\sim \gamma_0^2 \kperp^{-3},
\eeq
where $\gamma_0=(\kf^2\eps_0)^{1/3}$ is the rate of strain (the same for 
all modes), $\eps_0=\la(\vdperp\Phi_0)\cdot(\vdperp F_0)\ra$ the mean power injected into 
the $\kpar=0$ modes and $\kf$ the perpendicular wave number of 
the forcing \footnote{In a finite box, the energy 
will accumulate at the box wave number $\kbox$ and so 
$\gamma_0=\kbox(\eps_0 t)^{1/2}$.}. The spectra of the forced Alfv\'en waves 
and the $\kpar=0$ magnetic fluctuations follow from \eqref{adv_AW} with 
$\zo^\pm=\Phi_0$ and \eqref{amp_zmodes} with $\zo^\pm=\Psi_0$, respectively:
\begin{align}
&\Ekf^\pm(\kperp)\sim\kperp[\zkf^\pm(\kperp)]^2\sim\frac{\eps_1}{\gamma_0}\,\kperp^{-1},\\
&\Mo(\kperp)\sim\kperp[\Psi_0(\kperp)]^2\sim\lt(\frac{\eps_1}{\omA\gamma_0}\rt)^2\kperp,
\end{align}
where $\eps_1$ is the mean power injected into the $\kpar=\kparf$ Alfv\'en waves
(assumed balanced). 
This situation is illustrated in \figref{fig_invcasc}. 
The intersection wave number between $\Eo$ and $\Ekf$ is 
$\ko=(\gamma_0^3/\eps_1)^{1/2}\sim\kf$ if $\eps_0\sim\eps_1$. 
The intersection between $\Eo$ and $\Mo$ is at  
$\km=(\gamma_0^2\omA/\eps_1)^{1/2}$. 
For $\kperp>\km$, the rate of strain $\gamma_0$ of the hydrodynamic 
condensate is smaller than the rate of strain associated with the 
$\kpar=0$ modes induced by the oscillatory coupling to the Alfv\'en waves 
($\gamma_0\tnlo<1$) and so the turbulence reverts to the regime described earlier 
for the case of no forcing of the $\kpar=0$ modes.  

\paragraph{Other regimes.} Possibilities proliferate if we allow 
imbalanced energy injection and/or let the $\kpar=0$ modes 
be forced magnetically as well as hydrodynamically. The majority of the resulting 
regimes are probably not physically realizable, so we will not pursue this 
line of inquiry further, except for the following observation. If 
$\Psi_0$ is forced, this direct injection of energy into 2D magnetic fluctuations 
will in general trump the oscillatory coupling to Alfv\'en waves (the right-hand side 
of \eqref{Psi0_eq}). Thus, the 2D condensate decouples fully from the latter.
Magnetic field of the $\kpar=0$ modes is now 
dynamically significant, so there will be a 2D inverse cascade of the 
variance of the magnetic flux function, $|\Psi_0|^2$, and a direct cascade 
of the two Elsasser energies $|\dperp\zo^\pm|^2$. 
The inverse cascade of $|\Psi_0|^2$ to $\kperp<\kf$ 
produces a $\kperp^{-1/3}$ spectrum of magnetic energy and possibly 
a shallower spectrum of the kinetic energy \cite{Biskamp_Bremer}.
This may well be the physical mechanism responsible for the relative preponderance 
of magnetic energy over kinetic at large scales in weak MHD turbulence 
\cite{Wang_Boldyrev_Perez}. 
The direct cascade is numerically known to give rise to 
$\kperp^{-3/2}$ spectra \cite{Biskamp_Welter,Biskamp_Schwarz,Ng_etal_2D}\footnote{A possible 
explanation for this in the form of an adaptation of the dynamical alignment 
argument \cite{Boldyrev_align} to 2D is found at the end of Sec.~4 of 
\cite{Nazarenko_Schekochihin}.}. The Alfv\'en waves ($\kpar=\kparf$) will 
be passively mixed by this strong 2D MHD turbulence. However, by the causality argument 
given in the Introduction, the latter is, in fact, likely to be unstable, 
develop parallel decorrelations and become strong, 3D and critically balanced. 


\begin{acknowledgments} 
This work was supported by STFC (AAS and TAY) 
and the Leverhulme Trust Network for Magnetised Plasma Turbulence.
Much of it was completed during the Programme ``Frontiers in Dynamo Theory'' 
at Institut H.\ Poincar\'e, Paris (March-April 2009), whose hospitality 
AAS and SVN gratefully acknowledge. 
\end{acknowledgments}

\end{document}